\documentclass[11pt,a4paper]{article}
\pdfoutput=1
\usepackage{jcappub}

\usepackage{amssymb} %mathematical stuff
\usepackage{mathrsfs} %simboli matematici
\usepackage{multirow} %per accorpare righe sulle tabelle
\usepackage{subfig} %per mettere più immagini in un unico ambiente
\usepackage{bm} %grassetto per simboli matematici
\usepackage{graphicx}
\usepackage{amsmath,color}
\usepackage{amsfonts}
\newcommand{\mincir}{\raise
  -2.truept\hbox{\rlap{\hbox{$\sim$}}\raise5.truept \hbox{$<$}\ }}
\newcommand{\magcir}{\raise
  -2.truept\hbox{\rlap{\hbox{$\sim$}}\raise5.truept \hbox{$>$}\ }}
\newcommand{\siml}{\raise
  -2.truept\hbox{\rlap{\hbox{$\sim$}}\raise5.truept \hbox{$<$}\ }}
\newcommand{\simg}{\raise
  -2.truept\hbox{\rlap{\hbox{$\sim$}}\raise5.truept \hbox{$>$}\ }}

\newcommand{\eff}{\text{eff}}

\begin{document}

\title{Neutrino constraints: what large-scale structure~and~CMB~data~are~telling~us?}

\author[a,b]{Matteo Costanzi,}
\author[a,b]{Barbara Sartoris}
\author[c,b]{Matteo Viel,}
\author[a,b,c]{Stefano Borgani}

\affiliation[a]{Astronomy Unit, Department of Physics, University of Trieste, \\
  via Tiepolo 11, I-34143 Trieste, Italy}
\affiliation[b]{INFN-National Institute for Nuclear Physics,\\ via Valerio 2, I-34127 Trieste, Italy}
\affiliation[c]{INAF-Osservatorio Astronomico di Trieste,\\ via Tiepolo 11, I-34143 Trieste, Italy}

\emailAdd{costanzi@oats.inaf.it}
\emailAdd{sartoris@oats.inaf.it}
\emailAdd{viel@oats.inaf.it}
\emailAdd{borgani@oats.inaf.it}

\abstract{We discuss the reliability of neutrino mass constraints, either active 
or sterile, from the combination of different low redshift Universe probes with measurements of CMB anisotropies.
In our analyses we consider WMAP 9-year or Planck Cosmic Microwave Background (CMB) data 
in combination with Baryonic Acoustic Oscillations (BAO) measurements from BOSS DR11,
galaxy shear measurements from CFHTLenS, SDSS Ly-$\alpha$ 
forest constraints and galaxy cluster mass function from \textit{Chandra} observations.
At odds with recent similar studies, to avoid model dependence 
of the constraints we perform a full likelihood analysis for all the datasets employed.
As for the cluster data analysis we rely on to the most recent calibration of
massive neutrino effects in the halo mass function
and we explore the impact of the uncertainty 
in the mass bias and re-calibration of the halo mass function due to baryonic feedback processes on cosmological parameters.
We find that none of the low redshift probes alone provide 
evidence for massive neutrino in combination with CMB measurements,
while a larger than $2\sigma$ detection of non zero neutrino mass, either active or sterile,
is achieved combining cluster or shear data with CMB and BAO measurements.
Yet, the significance of the detection exceeds $3\sigma$ if we combine all four datasets.
For a three active neutrino scenario, from the joint analysis
of CMB, BAO, shear and cluster data including the uncertainty in the mass bias we obtain 
$\sum m_\nu =0.29^{+0.18}_{-0.21}$ eV and $\sum m_\nu =0.22^{+0.17}_{-0.18}$ eV ($95\%$CL)
using WMAP9 or Planck as CMB dataset, respectively.
The preference for massive neutrino is even larger in the sterile neutrino scenario,
for which we get $m_s^{\eff}=0.44^{+0.28}_{-0.26}$ eV and $\Delta N_\eff=0.78^{+0.60}_{-0.59}$ ($95\%$CL)
from the joint analysis of Planck, BAO, shear and cluster datasets.
For this data combination the vanilla $\Lambda$CDM model is rejected at more than $3\sigma$ and 
a sterile neutrino mass as motivated by accelerator anomaly is within the $2\sigma$ errors.  
Conversely, the Ly-$\alpha$ data favour vanishing neutrino masses and from 
the data combination Planck+BAO+Ly-$\alpha$ we get the tight upper limits 
$\sum m_\nu <0.14$ eV and $m_s^{\eff}<0.22$ eV -- $\Delta N_\eff<1.11$ ($95\%$CL) for the active and sterile neutrino model, respectively.
Finally, results from the full data combination reflect the tension between the $\sigma_8$ constraints obtained from cluster and shear data
and that inferred from Ly-$\alpha$ forest measurements; in the active neutrino scenario for both CMB datasets employed, the full data combination
yields only an upper limits on $\sum m_\nu$, while assuming an extra sterile neutrino we still get preference 
for non-vanishing mass, $m_s^{\eff}=0.26^{+0.22}_{-0.24}$ eV, and dark contribution to the radiation content, $\Delta N_\eff=0.82\pm0.55$.}

\maketitle

%%%%%%%%%%%%%%%%%%%%%% INTRODUCTION %%%%%%%%%%%%%%%%%%
\section{Introduction}\label{sec_int}
Results from Planck measurements of the cosmic microwave background (CMB) temperature anisotropies have yielded sub-percent
level constraints on the cosmological parameters of the concordance $\Lambda$CDM model~\citep{Planck_cosm_parameters}. The Planck CMB data
by themselves appear to be well described by the six standard $\Lambda$CDM parameters, and show no preference for extended models.
However, in the framework of the $\Lambda$CDM model, several probes of the low redshift Universe exhibit tension with the Planck results.
In particular, Planck finds a larger and more precise value of the matter density at recombination than previous CMB data; this results in a lower
value for the current expansion rate $H_0$, and a higher value of the matter density fluctuations $\sigma_8$.
These changes lead to a $\sim 2-3\sigma$ tension with direct measurements of $H_0$~\citep{Riess2011} and
$\sigma_8$ measurements from galaxy shear power spectrum~\citep{CFHTLenS2013}, galaxy-galaxy lensing~\citep{2013MNRAS.432.1544M}, redshift space distortion
(RSD)~\citep{Beutler2013} and clusters abundance~\citep{2009ApJ...692.1060V,2010ApJ...708..645R,2011ARA&A..49..409A,PlanckSZ2013}.
Meanwhile, agreement with distance measurements from Baryon Acoustic Oscillations (BAO) suggests that the discrepancy cannot be
resolved involving exotic dark energy models or curvature which modify the recent expansion history. 

Beside unresolved systematic effects neutrinos can offer a possible means to relieve this tension. 
Sterile neutrinos change the expansion rate at recombination and hence the calibration of the standard ruler with
which CMB and BAO observations infer distances, while massive neutrino, either sterile or active, suppress small
scale clustering at late time, relieving the tension with $\sigma_8$ measurements~\citep[see][for reviews]{2006PhR...429..307L,2011ARNPS..61...69W,2013neco.book.....L}.  
Massive neutrinos are an appealing solution since oscillation experiments with solar and atmospheric neutrinos have
already provided evidence for their mass, with room for extra sterile species, supported by anomalies in short baseline 
and reactor neutrino experiments~\citep{PhysRevD.86.010001,2012JHEP...12..123G,2013ARNPS..63...45C,2012arXiv1204.5379A}.
A number of studies have carried out joint analyses of various data combinations finding that a neutrino mass of $0.3-0.4 ~{\rm eV}$ 
provides a better fit to CMB data with low redshift Universe measurements than the vanilla $\Lambda$CDM 
model~\citep{2013AstL...39..357B,Wyman2013,Battye2013,Hamann2013,2013PhRvD..87l5034A,Beutler2014,2014arXiv1403.4852G,2014JCAP...06..031A}, 
although these conclusions are not universally accepted~\citep{2013JCAP...04..036F,2013JCAP...09..013V,2014arXiv1404.5950L}. 
While, none of these low redshift datasets combined individually with CMB measurements provide strong evidence for non-zero neutrino masses, the hint
for neutrino mass is driven mainly by low redshift growth of structure constraints (e.g. from shear and RSD measurements or cluster number counts).

In particular, galaxy clusters offer a powerful complementary probe to the CMB and
geometric probes as BAO thanks to the tight constraints provided
on the so called cluster normalization condition, $\sigma_8 \Omega_{\rm m}^\gamma$
%, where $\gamma\sim 0.5$, $\sigma_8$ is the r.m.s. linear matter fluctuations in spheres of $8h^{-1}$Mpc~\footnote{Here $h$ is defined such that $H_0=100 h$km s$^{-1}$ Mpc$^{-1}$}
%and $\Omega_{\rm m}$ is the current total matter density 
~\citep[see][for reviews in
cluster cosmology]{2011ARA&A..49..409A,2012ARA&A..50..353K}. X-ray~\citep{2010MNRAS.406.1759M,2009ApJ...692.1060V},
Sunyaev-Zel'Dovich~\citep{SPT2013,ACT2013,PlanckSZ2013} and optical~\citep{2010ApJ...708..645R} cluster surveys yield consistent
results favouring a value for the cluster normalization condition lower than the value derived from Planck data.
This tension between Planck and cluster data and the combination with BAO results could be taken as an evidence for non-vanishing neutrino masses.
However, the robustness of such constraints from cluster number counts depend on our capability to recover cluster masses from proxies
and to have precise theoretical predictions for the spatial number density of halos (the halo mass function, HMF)~\cite{Reed2013, Crocce_2010, Watson}.
Thus it is worth to investigate possible sources of systematic errors in cluster data which could lead to misinterpretation of the results
and to assess which combination of low-redshift datasets with CMB data prefers a non-zero neutrino mass within a given cosmological model.

In this work we derive constraints on neutrino properties combining WMAP9~\citep{2012arXiv1212.5226H}
 or Planck~\citep{Planck_cosm_parameters} CMB data with several probes of the local Universe: 
BOSS DR11 BAO scale~\citep{BOSS_DR11} and CFTHLenS shear~\citep{CFHTLenS2013} measurements,
SDSS Ly-$\alpha$ forest power spectrum constraints~\citep{2005ApJ...635..761M}
 and cluster mass function from \textit{Chandra} observations~\citep{2009ApJ...692.1033V}.
We consider two possible extensions of the standard $\Lambda$CDM model:
 a scenario with three degenerate active neutrinos, or a sterile massive neutrino model with
three active neutrinos distributed as in the minimal normal hierarchy scenario.
So far, many works~\citep[e.g.][]{Battye2013,Hamann2013,Beutler2014,2014arXiv1403.4852G} derived neutrino mass constraints
including cluster and/or shear constraints obtained within a $\Lambda$CDM model.  
In order to avoid misleading results due to model dependence of the constraints we perform a full likelihood analysis for all the datasets employed in this work, without
doing a sampling of the posterior probability (as in e.g. \citep{Wyman2013}).
In particular for the clusters data analyses we take into
account the effect of possible bias in the mass estimation and adopt different prescriptions for the HMF.
As for the latter, we consider the correction to the HMF proposed by~\citep{Ichiki-Takada}
 and~\citep{Castorina_2013} for cosmology with massive neutrinos. 
Moreover, we investigate how the different calibration of the HMF due to baryonic
feedback processes presented in~\citep{Cui2014} affects the cosmological constraints.

The paper is organized as follows. Section~\ref{sec:meth} describes
the cosmological models and datasets used in this work.
In section~\ref{sec:res} we present and discuss our results .
Finally, we draw the main conclusions in section~\ref{sec:conc}.

%%%%%%%%%%%%%%%%%%%%%%% Data and Methods %%%%%%%%%%%%%%%%%%
\section{Cosmological data analysis}\label{sec:meth}
\subsection{Models}
The baseline scenario analysed in this work is a $\Lambda$CDM
model with three degenerate massive neutrinos, defined by the parameters:
\begin{equation}
\{ \Omega_{\rm c}h^2, \Omega_{\rm b}h^2, \Theta_{\rm s}, \tau,n_{\rm s}, \log(10^{10}\,A_{\rm s}), \sum m_\nu \},
\end{equation}
with $\Omega_{\rm c}h^2$ and $\Omega_{\rm b}h^2$ being the physical cold dark matter and baryon energy densities,
$\Theta_{\rm s}$ the ratio between the sound horizon and the angular diameter distance at decoupling, $\tau$ the Thomson
optical depth at reionization, $n_{\rm s}$ the scalar spectral index, $A_{\rm s}$ the amplitude of the primordial power
spectrum and $\sum m_\nu$ the total neutrino mass. Note that given the current precision of cosmological constraints from
available data, the effect of mass splitting is negligible and the degenerate model can 
be assumed without loss of generality~\citep[see e.g.][]{2010JCAP...05..035J}.
We then consider a scenario with a massive sterile neutrino component
which has been suggested as a possible solution for the reactor~\cite{2011PhRvD..83g3006M}, 
Gallium~\cite{2006PhRvC..73d5805A,2012PhRvD..86k3014G} and accelerator~\cite{Aguilar:2001ty} anomalies in neutrino oscillation experiments.
Reactor and Gallium experiments prefer a new mass squared difference of $\Delta m^2 \gtrsim 1$ eV$^2$, while various
accelerator experiments constrain $\Delta m^2$ to be $\sim 0.5$ eV$^2$ (see~\cite{2012arXiv1204.5379A} and reference therein).
For this model we assume one massive active neutrino with $\sum m_\nu=0.06$ eV (the minimum mass allowed by neutrino oscillation experiments)
and we introduce two parameters to describe the extra hot relic component: the effective number of extra relativistic degree of
freedom $\Delta N_{\rm eff}$ and the effective sterile neutrino mass $m_{\rm s}^{\rm eff}$.
The former parametrizes any contribution to the radiation energy content ($\rho_{\rm r}$) besides photons
in the radiation dominated era through the formula:
\begin{equation}\label{eqn:neff2}
 \rho_{\rm r} = \left[ 1 + \frac{7}{8} \left(\frac{4}{11}\right)^{4/3} N_{\rm eff}\right] \rho_\gamma .
\end{equation}
In the standard model $N_{\rm eff}=3.046$ accounts for the three active neutrino species; thus $\Delta N_{\rm eff}=N_{\rm eff} - 3.046 > 0$
indicates new physics beyond the standard model: an extra thermalised light fermion
would contribute $\Delta N_{\eff}=1$, but more generally a non-integer $\Delta N_{\eff}$ value could arise from different physical phenomena, such as
lepton asymmetries~\citep{2012JCAP...07..025H}, partial thermalisation of new fermions~\citep{2009JCAP...01..036M}, particle
decay~\citep{2005JHEP...09..048P}, non-thermal production of dark matter~\citep{2012PhRvD..85f3513H,2014EPJC...74.2797K}, gravity
waves~\citep{2006PhRvL..97b1301S} or early dark energy~\citep{2011PhRvD..83l3504C}. The large values of the mass squared difference and mixing angles invoked to resolve reactor,
 Gallium ($\sin^2 2\Theta \gtrsim 0.1$) and accelerator ($\sin^2 2\Theta \sim 5 \times 10^{-3}$) anomalies suggest 
a fully thermalisation of the sterile neutrino in the early Universe~\cite{Langacker:1989sv}, and thus a contribution of $\Delta N_\eff=1$ to
the dark radiation.
The parameter $m_{\rm s}^{\rm eff}$, in the case of
thermally-distributed sterile neutrino, is related to the physical mass $m_{\rm s}$ via
\begin{equation}\label{eqn:meff}
 m_{\rm s}^{\rm eff} = (T_{\rm s}/T_{\nu})^3 m_{\rm s} = (\Delta N_{\eff})^{3/4} m_{\rm s}
\end{equation}
where $T_{\rm s}$ and $T_{\nu}$ represent the current temperature of the 
sterile and active neutrinos, respectively. Alternatively, if the sterile
neutrino is distributed proportionally to the active species due to oscillations
 the physical mass can be expressed as $m_{\rm s}^{\rm eff} = (\Delta N_{\eff}) m_{\rm s}$, which corresponds to the
Dodelson-Widrow scenario~\citep{1994PhRvL..72...17D}. In both cases for a fully 
thermalised sterile neutrino, $\Delta N_\eff=1$, one gets $m_{\rm s}^{\rm eff}=m_{\rm s}$.
In our analysis we adopt the prior $m_{\rm s}^{\rm eff}/(\Delta N_{\eff})^{3/4} < 10$ eV
 to avoid a degeneracy between very massive neutrinos and cold dark matter.

\subsection{Data and analysis}
We infer posterior probability distributions by means of the Monte Carlo Markov Chain technique using the publicly available code
{\tt CosmoMC}\footnote{http://cosmologist.info/cosmomc/}~\citep{Lewis2002} for various combinations of the following datasets: 

\textit{CMB} -- We consider CMB temperature and polarization measurements from 9-year WMAP data release (hereafter WMAP9)~\citep{2012arXiv1212.5226H}
or, alternatively, temperature power spectrum from the Planck satellite~\citep{Planck_cosm_parameters} combined with large-scale TE- and EE-polarization
power spectra from WMAP9 (hereafter Planck). These datasets are analysed using
the likelihood functions provided by the Planck collaboration~\citep{PlanckXV},
and publicly available at Planck Legacy Archive\footnote{http://pla.esac.esa.int/pla/aio/planckProducts.html}
and marginalizing over the foreground nuisance parameters. The helium abundance is computed as a function of
 $\Omega_{\rm b}h^2$ and $N_{\rm eff}$, following the Big Bang Nucleosynthesis theoretical predictions. In the
 Planck analysis we fix the lensing spectrum normalization parameter to $A_{\rm L}=1$, if not otherwise stated.
The WMAP9 dataset is not sensitive to the gravitational lensing signal since its effects can be detected only
at large multipoles.

\textit{BAO} -- We include the most recent and accurate measurements of the BAO scales from BOSS Data Release 11~\citep{BOSS_DR11}.
Exploiting a sample of nearly one million galaxies observed over $8500$ square degree between redshift $0.2<z<0.7$, DR11 results provides
percent level constraints on the peak position of the spherically averaged galaxy correlation function at redshift $z=0.32$ and $z=0.57$.
The likelihood function associated to this dataset is estimated using the likelihood code distributed with the {\tt CosmoMC} package.

\textit{Shear} -- We use the 6-bin tomography angular galaxy shear power spectra data from the
CFTHLenS survey~\citep{CFHTLenS2013}\footnote{http://cfhtlens.org/astronomers/cosmological-data-products}. 
The survey spans over 154 square degrees in five optical bands, with shear and photometric redshift measurements
for a galaxy sample with a median redshift of $z=0.70$.
Constraints from this datasets are derived using a modified version of the {\tt CosmoMC}
 module~\footnote{http://www.astro.caltech.edu/~rjm/cosmos/cosmomc/}
for the weak lensing COSMOMS 3D data~\citep{2007JCAP...11..008L}. The code has been substantially modified in order to reproduce
the analysis described in~\citep{CFHTLenS2013} which makes use of 21 sets of cosmic shear correlation functions associated to
6 redshift bins, each spanning the angular range of $1.5-35$ arcmin, to extract cosmological information.
As in~\citep{CFHTLenS2013}, we also include in the module the model for the intrinsic alignment treatment
developed by~\citep{2007NJPh....9..444B}, which accounts for both intrinsic alignment of physically nearby
galaxies and the shear-shape correlation for galaxies separated by large physical distances along the line of sight. 
This model, which is based on a fitting approach, has the advantage of needing only one additional 
nuisance parameter, marginalized over in the analysis, to predict both the intrinsic alignment contributions to the shear correlation functions.
We verified that our module reproduces well the results presented in~\citep{CFHTLenS2013} for a $\Lambda$CDM model.
  
\textit{Ly-$\alpha$} -- We rely on the SDSS Ly-$\alpha$ forest data from~\citep{2005ApJ...635..761M} to constrain the amplitude,
slope and curvature of the linear matter power spectrum at scale $k=0.009\,$s km$^{-1}$ and redshift $z=3$. We combine this
dataset by implementing the Ly-$\alpha$ likelihood code distributed with the {\tt CosmoMC}
package. The module has been updated to work with the new version of {\tt CosmoMC} and it has been implemented with the patch
written by A. Slosar~\footnote{http://www.slosar.com/aslosar/lya.html} in order to support extended model analysis.
Note that this data set does not include the most recent BOSS data of ~\citep{palanque} that will soon provide an updated
value of the upper limits obtained in \citep{2006JCAP...10..014S}, by using a new technique to sample the parameter space \citep{borde} and 
hydro simulations that incorporate massive neutrinos \citep{rossi}.
 
\textit{Clusters} -- Constraints from galaxy clusters are obtained exploiting the CCCP catalogue presented in~\citep{2009ApJ...692.1033V}. 
The catalogue consists of X-ray \textit{Chandra} observations of $37$ clusters with $\langle z\rangle=0.55$ derived from the 400 deg$^2$ \textit{ROSAT}
survey and $49$ brightest $z\approx 0.05$ clusters detected in the \textit{ROSAT} All-Sky Survey, which provide a robust determination of the cluster mass function at low and high redshifts.
To derive cosmological constraints we developed our own module for {\tt CosmoMC} following the fitting procedure outlined in~\citep{2009ApJ...692.1033V}.
For the cluster masses we use $Y_x$ proxy mass estimations~\citep{2006ApJ...650..128K}, which allows us to implement
the X-ray luminosity-mass relation presented in~\citep{2009ApJ...692.1033V} needed to compute the survey volume as a function of the mass.
The theoretical abundance of massive halo is computed using the Tinker HMF~\citep{Tinker2008}, where the coefficient
of the fitting formula are obtained interpolating table 2 of~\citep{Tinker2008} for halos 
with $\Delta_{\rm mean}=\Delta_{\rm critical}/\Omega_{\rm m}=500/\Omega_{\rm m}$,
according to the cluster mass definition of~\citep{2009ApJ...692.1033V}.
We verified that our analysis reproduces accurately the results of~\citep{2012AstL...38..347B} for the combination of WMAP 7-year and CCCP cluster data.
To properly take into account the effects of massive neutrinos on the HMF calibration, we neglect the weakly clustering neutrino
component when calculating the halo mass, as suggested by many authors~\citep[e.g.][]{Brandbyge_halos,Marulli_2011,Paco_2013a}.
Moreover, following~\citep{Ichiki-Takada,Castorina_2013} the variance of the matter perturbations, required to predict the HMF,
is computed using only the cold dark matter and baryon linear power spectrum, in order to neglect
the suppression of the matter density fluctuations on scales smaller than the neutrino free-streaming length.
These corrections entail an increase of the HMF with respect
  to the previous calibration. This effect is larger
for larger neutrino masses and higher number of massive neutrino
species. In turn, the increase in the HMF affects the resulting
constraints on cosmological parameters, e.g.
by steepening the $\sigma_8-\Omega_{\rm m}$ degeneracy direction, thereby reducing the $\sigma_8$ mean value~\citep{Costanzi2013b}.
Besides these modifications to the original analysis of~\citep{2009ApJ...692.1033V} we test the effect of other sources of systematics.
To address the impact on clusters constraints due to baryonic feedback processes we implement the correction to the HMF proposed by~\citep{Cui2014};
the net effect of baryonic processes is to generate shallower density
profiles and a corresponding decrease of halo masses with respect to 
the dark matter only case used to fit the HMF~\citep[e.g.][]{Jenkins, Reed, Warren, Tinker2008, Crocce_2010}.
This effect is taken into account correcting the halo masses through a mass dependent fit to the halo mass variation induced by baryonic processes. 
Similar studies have been carried out by different
groups~\citep{2014MNRAS.439.2485C,2014MNRAS.442.2641V,2014arXiv1405.2921V,2014MNRAS.440.2290M}
that found consistent results (cf. also \citep{2014MNRAS.440.2290M}).

The main source of systematic errors is related to the uncertainty in
cluster mass measurements. For the catalogue used in this work cluster
masses have been inferred by using a scaling relation between total
mass and the product of hot intracluster gas mass and temperature, as
inferred from X-ray observations. This scaling relation has been
calibrated by resorting to X--ray hydrostatic mass measurements for
nearby galaxy clusters~\citep[see][for details]{2009ApJ...692.1033V}.
However, mass measurements from X--ray measurements suffer for
different sources of possible systematic biases, e.g. associated to
departures from spherical symmetry, to biases in X-ray measurement
of gas temperature \citep[e.g.][]{2008MNRAS.384.1567M,2012NJPh...14e5018R}, or to
violation of hydrostatic equilibrium due to the presence non-thermal
pressure support~\citep[e.g.][]{2010A&A...510A..76L,2014arXiv1401.7657S}.

To consider this uncertainty we introduce a nuisance bias parameter
defined as $M^{est}/M^{true}=B_M$, which is varied in the range
$[0.8:1.0]$ when included in the fit. This prior accounts for the
constraints on the mass bias from~\citep{Israel2014}
and~\citep{2014arXiv1405.7876D} obtained comparing \textit{Chandra}
X-ray masses with MMT/Megacam weak lensing masses of 8 clusters
between redshift $0.39-0.80$ and CLASH weak lensing masses of 20
clusters between redshift $0.2-0.5$, respectively. A different
  range for the mass bias, $B_M \simeq 0.7 \pm 0.1$, is suggested by
  the analysis by~\citep{2014arXiv1402.2670V} which however uses X-ray
  masses derived from XMM--Newton temperature measurements.
  Temperature measurements from XMM--Newton could be systematically
  lower with respect to those obtained from \textit{Chandra}
  observations, as discussed
  by~\citep{2010A&A...523A..22N,2014arXiv1404.7130S,2014arXiv1405.7876D},
  thus providing a larger mass bias.

%%%%%%%%%%%%%%%%%%%%% mass function %%%%%%%%%%%%%%%%%%%%%
\section{Results} 
\label{sec:res}
For each of the cases that we describe here below, we run four independent chains,
requiring the fulfilment of the Gelman \& Rubin~\cite{1992...Gelman}
criteria with $R-1\leqslant 0.03$ as convergence test. The best fit values are obtained
with the {\tt BOBYQA} maximisation routine provided in {\tt CosmoMC}. If not otherwise stated,
errors and upper limits reported in the text have to be intended at $95\%$ confidence level.

\begin{figure}
\centering
\includegraphics[width=0.48\textwidth]{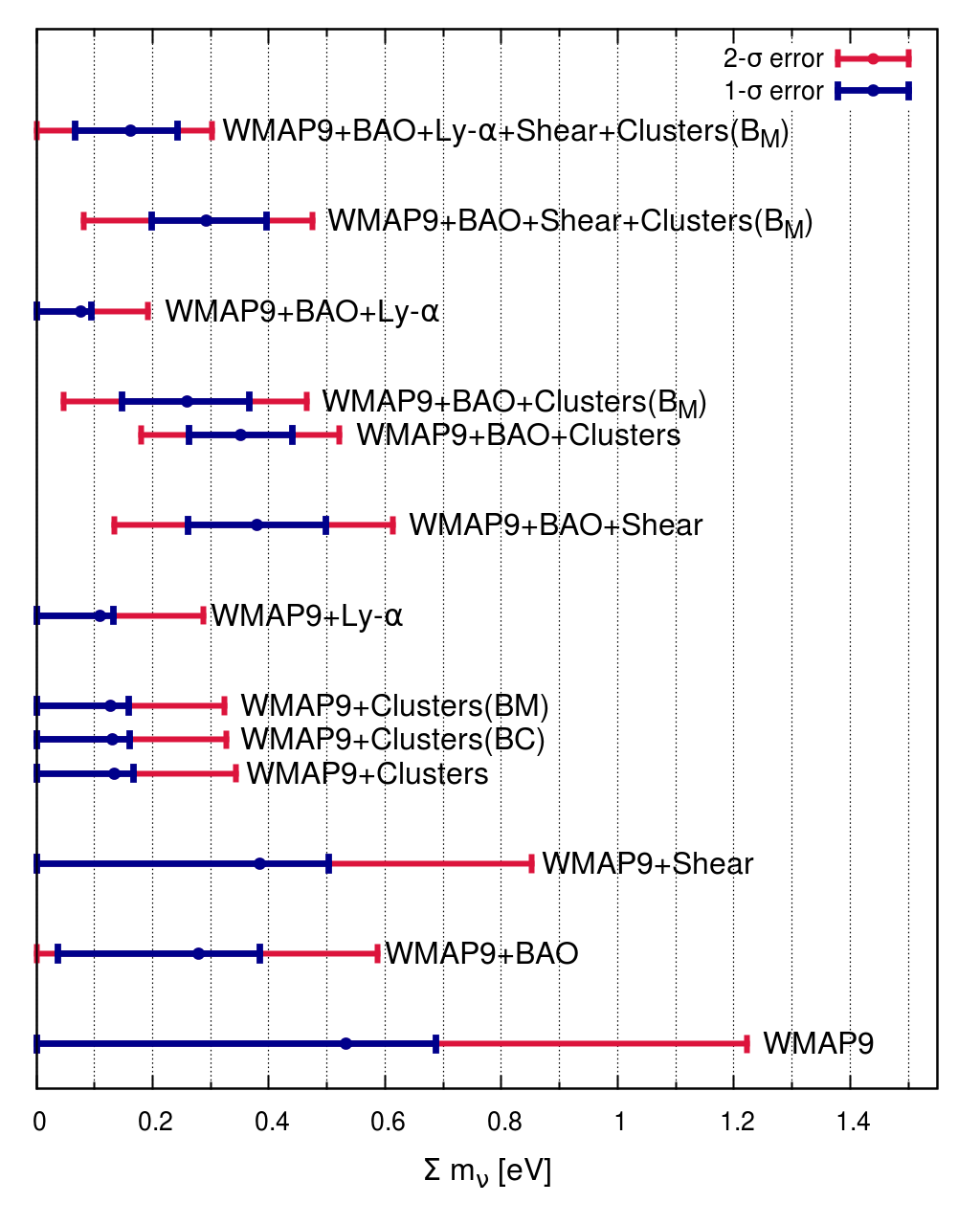}
\includegraphics[width=0.48\textwidth]{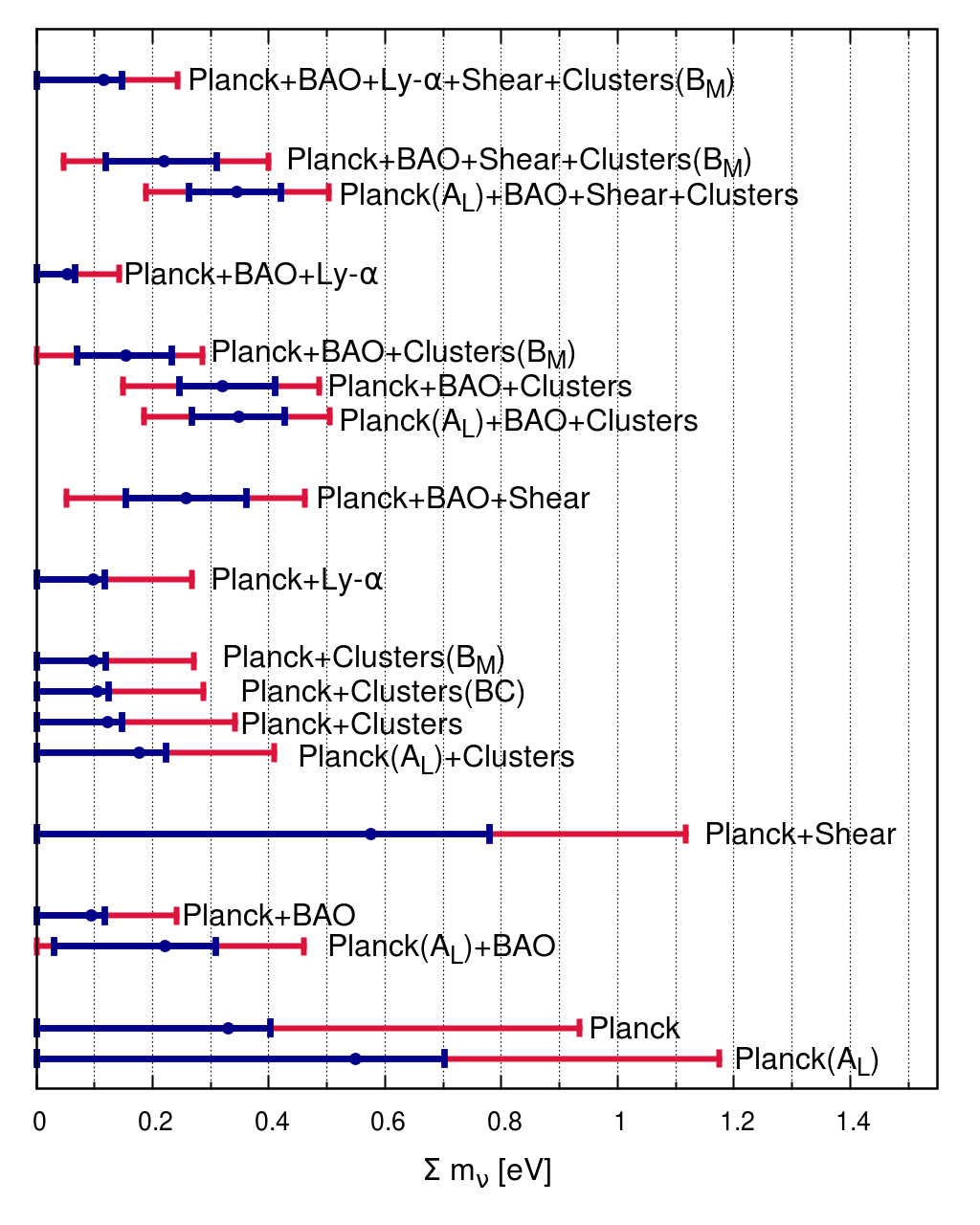}
\caption{Summary of the $1\sigma$ and $2\sigma$ errors on $\sum m_\nu$ obtained from the dataset combinations discussed in section~\ref{sub:mnu}
within a $\Lambda$CDM+$\sum m_\nu$ model.}
\label{fig:mnu_error}
\end{figure}

\begin{table}
    \centering
    %\scriptsize
    \footnotesize
    %\small
    \caption{Constraints on $\Omega_{\rm m}$, $\sigma_8$ and $\sum m_\nu$ for a $\Lambda$CDM+$\sum m_\nu$ model combining different datasets.
      Errors are reported at $68\%$ confidence level for $\sigma_8$ and $\Omega_{\rm m}$, and both $68\%$ and $95\%$ confidence level for $\sum m_\nu$.
      Notations included in parenthesis denote modifications to the standard setting: (BC) stands for the baryon correction to the HMF,
      while ($B_M$) and ($A_{\rm L}$) indicate analyses with the bias or lensing signal parameter marginalized out.}
    \label{tab:mnu}
    %\vspace{1mm}
   \begin{tabular}{lcccc}
    \hline 
	Dataset			&	$\Omega_{\rm m}$	&	$\sigma_8$	&	\multicolumn{2}{c}{$\sum m_\nu$[eV]}	\\
				&				&			&	$68\%$CL	&	$95\%$CL	\\
    \hline \vspace{-3mm}\\
	WMAP9			&$0.347^{+0.043}_{-0.079}$& $0.714^{+0.091}_{-0.068}$	& 	$<0.68$		& 	$<1.22$		\vspace{0.5mm} \\
	WMAP9+Cluster		&$0.264^{+0.010}_{-0.017}$& $0.780^{+0.027}_{-0.016}$	& 	$<0.17$		& 	$<0.34$		\vspace{0.5mm} \\
      WMAP9+Cluster($B_M$)	&$0.275^{+0.014}_{-0.020}$& $0.793^{+0.029}_{-0.019}$	& 	$<0.16$		& 	$<0.33$		\vspace{0.5mm} \\ 
	WMAP9+Cluster(BC)	&$0.271^{+0.011}_{-0.017}$& $0.789^{+0.026}_{-0.017}$	& 	$<0.16$		& 	$<0.33$		\vspace{0.5mm} \\
	WMAP9+BAO		&$0.304^{+0.009}_{-0.011}$& $0.759^{+0.062}_{-0.043}$	&$0.28^{+0.11}_{-0.24}$	& 	$<0.59$		\vspace{0.5mm} \\
	WMAP9+Shear		&$0.305^{+0.029}_{-0.055}$& $0.726^{+0.061}_{-0.046}$	& 	$<0.50$		& 	$<0.85$		\vspace{0.5mm} \\
	WMAP9+Ly-$\alpha$	&$0.320^{+0.026}_{-0.033}$& $0.830^{+0.025}_{-0.021}$	& 	$<0.13$		& 	$<0.29$		\vspace{0.5mm} \\
	WMAP9+BAO+Cluster	&$0.298^{+0.009}_{-0.009}$& $0.735^{+0.015}_{-0.032}$	&$0.35^{+0.09}_{-0.09}$	&$0.35^{+0.17}_{-0.17}$	\vspace{0.5mm} \\
  WMAP9+BAO+Cluster($B_M$)	&$0.298^{+0.009}_{-0.009}$& $0.765^{+0.024}_{-0.028}$	&$0.26^{+0.11}_{-0.11}$	&$0.26^{+0.20}_{-0.20}$	\vspace{0.5mm} \\
	WMAP9+BAO+Shear		&$0.303^{+0.010}_{-0.010}$& $0.724^{+0.028}_{-0.028}$	&$0.38^{+0.12}_{-0.12}$	&$0.38^{+0.23}_{-0.24}$	\vspace{0.5mm} \\
	WMAP9+BAO+Ly-$\alpha$	&$0.305^{+0.009}_{-0.009}$& $0.833^{+0.021}_{-0.019}$	&	$<0.09$		& 	$<0.19$		\vspace{0.5mm} \\
WMAP9+BAO+Cluster($B_M$)+Shear	&$0.297^{+0.010}_{-0.010}$&$0.752^{+0.017}_{-0.022}$	&$0.29^{+0.11}_{-0.9}$	&$0.29^{+0.18}_{-0.21}$	\vspace{0.5mm} \\
WMAP9+BAO+Ly-$\alpha$+Shear+Cluster($B_M$)&$0.289^{+0.008}_{-0.08}$& $0.787^{+0.019}_{-0.017}$&$0.16^{+0.08}_{-0.09}$&$<0.30$	\vspace{0.5mm} \\
   \hline \vspace{-3mm}\\
	Planck			&$0.355^{+0.025}_{-0.061}$& $0.775^{+0.077}_{-0.032}$	& 	$<0.40$		& 	$<0.93$		\vspace{0.5mm} \\
	Planck+Cluster		&$0.272^{+0.008}_{-0.018}$& $0.782^{+0.027}_{-0.013}$	& 	$<0.15$		& 	$<0.34$		\vspace{0.5mm} \\
	Planck+Cluster($B_M$)	&$0.287^{+0.011}_{-0.015}$& $0.802^{+0.025}_{-0.012}$	& 	$<0.12$		& 	$<0.27$		\vspace{0.5mm} \\
	Planck+Cluster(BC)	&$0.278^{+0.009}_{-0.017}$& $0.790^{+0.026}_{-0.013}$	& 	$<0.14$		& 	$<0.32$		\vspace{0.5mm} \\
	Planck+BAO		&$0.309^{+0.009}_{-0.009}$& $0.819^{+0.027}_{-0.016}$	&	$<0.12$		& 	$<0.24$		\vspace{0.5mm} \\
	Planck+Shear		&$0.358^{+0.051}_{-0.078}$& $0.708^{+0.093}_{-0.082}$	& 	$<0.78$		& 	$<1.12$		\vspace{0.5mm} \\
	Planck+Ly-$\alpha$	&$0.329^{+0.018}_{-0.024}$& $0.831^{+0.024}_{-0.015}$	& 	$<0.12$		& 	$<0.27$		\vspace{0.5mm} \\
	Planck+BAO+Cluster	&$0.300^{+0.010}_{-0.010}$& $0.741^{+0.015}_{-0.018}$	&$0.32^{+0.09}_{-0.07}$	&$0.32^{+0.17}_{-0.17}$	\vspace{0.5mm} \\
Planck+BAO+Cluster($B_M$)	&$0.300^{+0.007}_{-0.009}$& $0.791^{+0.020}_{-0.018}$	&$0.15^{+0.08}_{-0.08}$	&	$<0.28$		\vspace{0.5mm} \\
	Planck+BAO+Shear	&$0.306^{+0.010}_{-0.011}$& $0.763^{+0.025}_{-0.024}$	&$0.26^{+0.10}_{-0.10}$	&$0.26^{+0.20}_{-0.21}$	\vspace{0.5mm} \\
  Planck+BAO+Ly-$\alpha$	&$0.310^{+0.008}_{-0.008}$& $0.836^{+0.016}_{-0.014}$	&	$<0.07$		& 	$<0.14$		\vspace{0.5mm} \\
Planck+BAO+Cluster($B_M$)+Shear&$0.300^{+0.009}_{-0.011}$&$0.770^{+0.021}_{-0.021}$	&$0.22^{+0.09}_{-0.10}$&$0.22^{+0.17}_{-0.18}$	\vspace{0.5mm} \\
Planck+BAO+Ly-$\alpha$+Shear+Cluster($B_M$)&$0.293^{+0.009}_{-0.010}$& $0.798^{+0.018}_{-0.014}$&	$<0.15$	& 	$<0.24$		\vspace{0.5mm} \\
   \hline \vspace{-3mm}\\
Planck($A_{\rm L}$)	&$0.358^{+0.040}_{-0.067}$& $0.716^{+0.081}_{-0.066}$	& 	$<0.70$		& 	$<1.17$		\vspace{0.5mm} \\
Planck($A_{\rm L}$)+Cluster&$0.276^{+0.010}_{-0.022}$& $0.769^{+0.032}_{-0.018}$	& 	$<0.22$		& 	$<0.41$		\vspace{0.5mm} \\
Planck($A_{\rm L}$)+BAO		&$0.307^{+0.010}_{-0.010}$& $0.778^{+0.047}_{-0.033}$	&$0.22^{+0.09}_{-0.19}$	&	$<0.46$		\vspace{0.5mm} \\
Planck($A_{\rm L}$)+BAO+Cluster&$0.301^{+0.009}_{-0.009}$& $0.734^{+0.016}_{-0.015}$	&$0.35^{+0.08}_{-0.08}$	&$0.35^{+0.15}_{-0.16}$	\vspace{0.5mm} \\
Planck($A_{\rm L}$)+BAO+Shear+Cluster&$0.300^{+0.009}_{-0.010}$& $0.733^{+0.016}_{-0.014}$&$0.35^{+0.08}_{-0.08}$&$0.35^{+0.15}_{-0.16}$\vspace{0.5mm} \\
    \hline
    \end{tabular}
\end{table}

\begin{figure}
\centering
\includegraphics[width=0.722\textwidth]{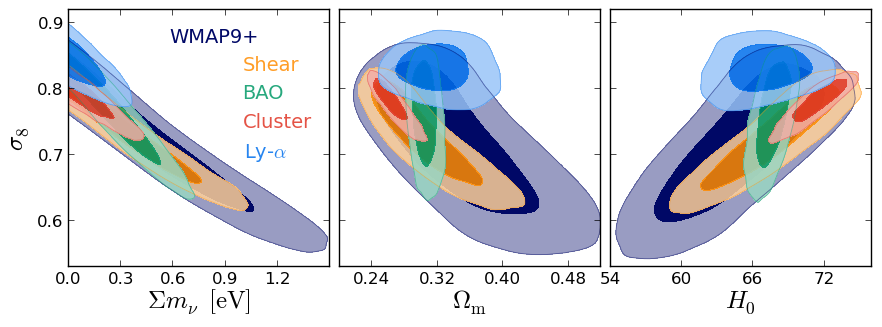}
\includegraphics[width=0.266\textwidth]{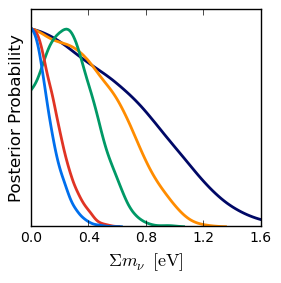}
\includegraphics[width=0.722\textwidth]{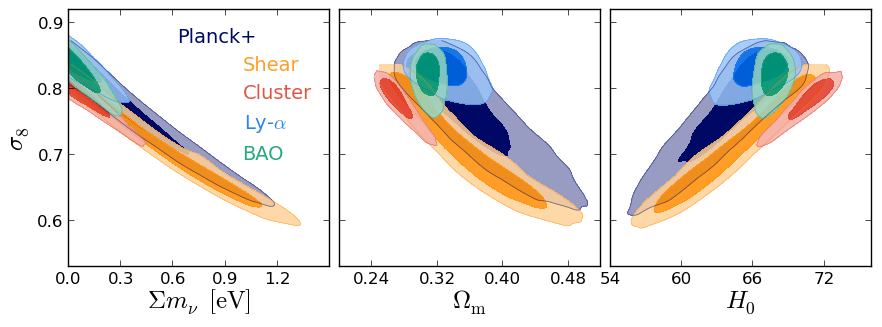}
\includegraphics[width=0.266\textwidth]{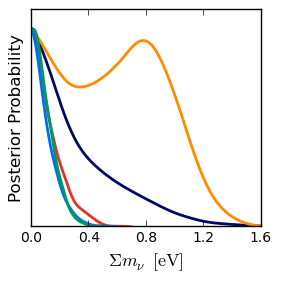}
\caption{\textit{Left} -Joint 68\% and 95\% CL contours in the $\sigma_8 - (\Omega_{\rm m},H_0,\sum m_\nu)$ planes
 for a $\Lambda$CDM+$\sum m_\nu$ model from the combination of WMAP9 (\textit{upper} panels) or Planck
 (\textit{lower} panels) data with different low redshift Universe probes.
\textit{Right} - Posterior probability distribution for $\sum m_\nu$ from the same data combination.}
\label{fig:CMB_single_mnu}
\end{figure}

\subsection{Massive neutrinos}\label{sub:mnu}
\begin{figure}
\centering
\includegraphics[width=0.722\textwidth]{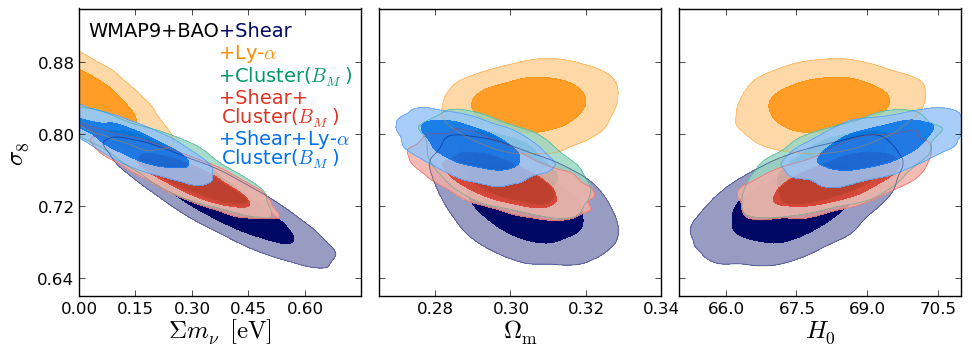}
\includegraphics[width=0.266\textwidth]{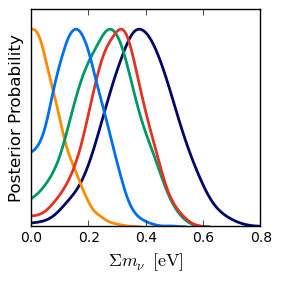}
\includegraphics[width=0.722\textwidth]{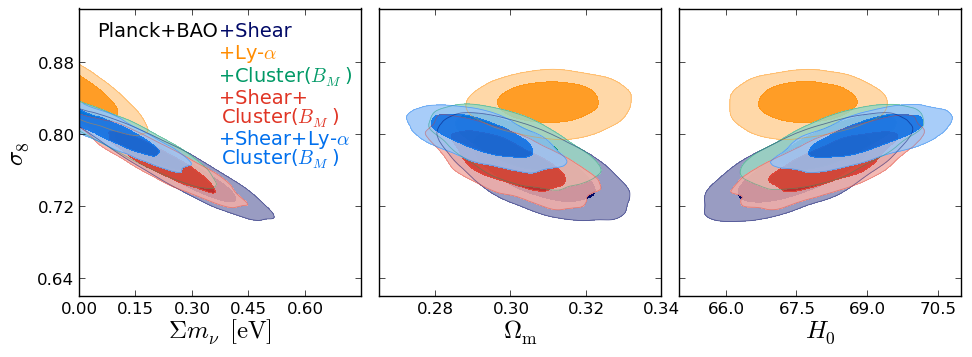}
\includegraphics[width=0.266\textwidth]{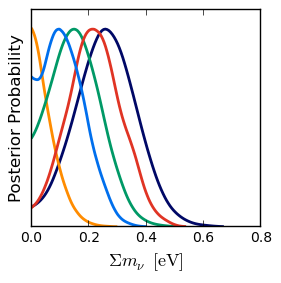}
\caption{\textit{Left} -Confidence contours at 68\% and 95\% CL in the $\sigma_8 - (\Omega_{\rm m},H_0,\sum m_\nu)$
 planes when combining  WMAP9 (\textit{upper} panels) or Planck (\textit{lower} panels) with 
many different probes of the low redshift Universe within a $\Lambda$CDM+$\sum m_\nu$ model.
\textit{Right} - Posterior probability distribution for $\sum m_\nu$ from the same datasets.}
\label{fig:CMB_all_mnu}
\end{figure}

%%%% Using WMAP9
We first turn our attention to the degenerate active neutrino case, whose results are summarized in Table~\ref{tab:mnu}
and Fig.~\ref{fig:mnu_error}.
To illustrate how different probes of the low redshift Universe
combined with CMB measurements constrain cosmological parameters we use them one by one,
before combining them together (see Fig.~\ref{fig:CMB_single_mnu} and Fig.~\ref{fig:CMB_all_mnu}, respectively). 

\begin{figure}
\centering
\includegraphics[width=0.325\textwidth]{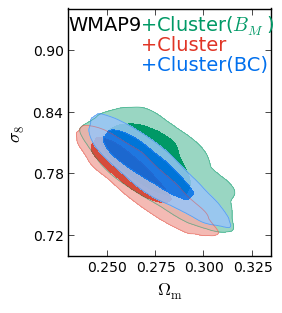}
\includegraphics[width=0.325\textwidth]{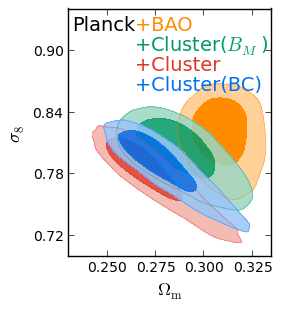}
\includegraphics[width=0.325\textwidth]{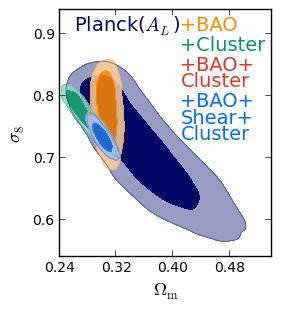}
\caption{\textit{Left-Middle panel} - Comparison of the confidence contours at 68\% and 95\% CL in
the $\sigma_8 - \Omega_{\rm m}$ plane within a $\Lambda$CDM+$\sum m_\nu$ model when combining  WMAP9
(\textit{left} panel) or Planck (\textit{middle} panel) with cluster data using different prescriptions:
the standard one (\textit{Cluster}), the baryon correction (\textit{Cluster(BC)}) or marginalizing over the bias (\textit{Cluster}($B_M$)).
In the \textit{middle} panel are also shown the confidence contours for the joint analysis Planck+BAO:
only when $B_M$ is allowed to vary the Planck+BAO and Planck+Cluster($B_M$) regions overlap. 
\textit{Right panel} - Joint 68\% and 95\% CL constraints on $\sigma_8 - \Omega_{\rm m}$ for 
different dataset combined with Planck with the $A_{\rm L}$-lensing signal marginalized out.}
\label{fig:mnu_variation}
\end{figure}

\textit{Combing with WMAP9} -- The upper panels of Fig.~\ref{fig:CMB_single_mnu} show constraints on the $\sigma_8 -
 (\sum m_\nu,\Omega_{\rm m},H_0)$ planes and the 1D likelihood distribution of $\sum m_\nu$ for several datasets combined with WMAP9 as CMB data. 
None of them exhibit tension with WMAP9 results nor evidence for non-zero neutrino mass.
The stronger constraint on the neutrino mass, $\sum m_\nu<0.29$ eV, comes
from the inclusion of Ly-$\alpha$ data, due to the high $\sigma_8$ value preferred by this dataset.
 Similar results (on $\sum m_\nu$) involve the inclusion of cluster data which shrinks and
shifts the $\sigma_8-\Omega_{\rm m}$ contours toward lower values requiring small values for the total neutrino mass, $\sum m_\nu < 0.34$ eV.
Repeating the analysis with a free mass bias parameter (Cluster($B_M$)) or taking into account
the baryon correction to the HMF (Cluster(BC)) slightly increases the $\sigma_8$ and $\Omega_{\rm m}$ values
(see left panel of Fig~\ref{fig:mnu_variation})
without significantly affecting the bound on $\sum m_\nu$ nor its best fit values.
In the latter case the errors on $\sigma_8$ and $\Omega_{\rm m}$ remain unchanged while the suppression of
the HMF with respect to the standard case causes the shift of the two parameters. Conversely, the inclusion
of $B_M$ in the fit relaxes the bounds on $\sigma_8$ and $\Omega_{\rm m}$ and shifts their contours
owing to the low value assumed by the bias, $B_M\sim0.9$.  
On the other hand, BAO data shows a mild preference for larger neutrino mass (see right panel of Fig~\ref{fig:CMB_single_mnu})
which displaces the neutrino bounds to higher values, $\sum m_\nu < 0.59$ eV, reason for that being the tight constraints 
on $\Omega_{\rm m}$ and the low $\sigma_8$ value allowed by this datasets combination.
Shear measurement, as cluster number counts, provides constraints on  
$\sigma_8 \Omega_{\rm m}^\gamma$ but with a poorer constraining power than clusters data and
with a degeneracy direction more similar to the one given by WMAP9 data; therefore the inclusion of this dataset
entails only a small improvement on neutrino mass constraints.

We start now to perform joint analyses of different probes of the low redshift Universe.
The results are presented in the upper panels of Fig~\ref{fig:CMB_all_mnu}.
Both the additions of cluster and shear datasets to the WMAP9+BAO joint analysis result in a larger than $2\sigma$
preference for massive neutrino yielding $\sum m_\nu=0.35 \pm 0.17$ eV and $\sum m_\nu=0.38^{+0.23}_{-0.24}$ eV, respectively.
Also when the bias parameter is marginalized out the combination WMAP9+BAO+Cluster($B_M$) shows a $2\sigma$ evidence
for non-zero neutrino masses, although with larger error bars and a lower mean value: $\sum m_\nu = 0.26 \pm 0.20$ eV.
The result can be understood as follows: the BAO scale measurements basically fix the matter density parameter thus
breaking the $\sigma_8-\Omega_{\rm m}$ degeneracy typical of cluster and shear constraints.
The tight constraints obtained for these two parameters along with the large value of $\Omega_{\rm m}$ (driven by the BAO data)
and small value of $\sigma_8$ (driven by cluster abundance or shear measurements) are compensated with a large value of $\sum m_\nu$.
At variance, Ly-$\alpha$ data prefers large value of the power spectrum normalization, and when joined with WMAP9+BAO data 
the large $\Omega_{\rm m}$ and $\sigma_8$ values inferred require small neutrino masses yielding the upper limit $\sum m_\nu <0.19$ eV.
The further inclusion of shear data in the WMAP9+BAO+Cluster($B_M$) analysis does not improve substantially the error on the neutrino
mass, but decreases $\sigma_8$ by $\sim 0.01$ thus favouring slightly larger neutrino masses and increasing to $3\sigma$ the significance 
of the mass detection.

Finally we jointly analyse the WMAP9, BAO, Ly-$\alpha$, shear and cluster data including $B_M$ into the fit obtaining $\sum m_\nu<0.30$ eV.
The addition of Ly-$\alpha$ forest measurements raises the power spectrum normalization by $\sim 0.04$ which causes a shift of the neutrino
mass toward lower values and reduces to $1\sigma$ the significance of the mass detection.  
We decided not to combine Ly-$\alpha$ and cluster data without marginalizing over the bias since the WMAP9+BAO+Ly-$\alpha$ and WMAP9+BAO+Cluster
data are already in tension by more than $2\sigma$ (see upper panels of Fig.~\ref{fig:CMB_all_mnu}). 

%%%% Using Planck
\textit{Combining with Planck} -- Now we repeat the same analysis but replacing the WMAP9 dataset with Planck.
The results are summarized in Table~\ref{tab:mnu} and shown in the lower panels of Fig~\ref{fig:CMB_single_mnu}
and Fig~\ref{fig:CMB_all_mnu}. Planck provides not only tighter cosmological constraints than WMAP9 data but it also 
prefers a larger matter density parameters, which in turn lowers the derived $H_0$ value and increases the power spectrum normalization. 
For datasets which are consistent with Planck measurements, i.e. BAO and Ly-$\alpha$, the combination with this CMB data yields lower
upper limits on the neutrino mass: $\sum m_\nu<0.27$ eV and $\sum m_\nu <0.24$ eV for Planck+Ly-$\alpha$ and Planck+BAO, respectively.
In this case the CMB+BAO combination does not show preference for large neutrino mass thanks to the larger value
and tighter constraints on $\sigma_8$ provided by Planck.
Conversely, the addition of shear or cluster data, which prefer lower $\sigma_8$, shifts the contours outside the region allowed
by Planck by $1\sigma$ and $2\sigma$, respectively. This indicates that the extension to massive neutrino is not sufficient
to bring the two datasets in agreement with Planck measurements. The shear measurements does not improve the constrains on $\sum m_\nu$,
while clusters number counts yields an upper limit of $0.34$ eV . Including in the cluster analysis the baryon correction to the HMF
increases by few percents the $\sigma_8$ and $\Omega_{\rm m}$ values improving the fit by $\Delta \chi^2\simeq2$,
but it is not sufficient to relieve the tension between the two datasets.
Allowing the bias to vary causes the contours to move towards the region allowed by Planck bringing the datasets in better agreement at 
the expense of a large mass bias, $B_M \sim0.8$. In this case the best fit $\chi^2$ is reduced by $\sim 9$ 
with respect to the standard Planck+Cluster analysis and, as expected for consistent datasets, the errors shrink
giving an upper limit of $\sum m_\nu<0.27$ eV.

As above we start now to combine different probes of the low redshift Universe at the same time.
The main results are shown in the lower panels of Fig.~\ref{fig:CMB_all_mnu}. Similar to the previous results 
the inclusion of cluster or shear datasets in the Planck+BAO joint analysis results in a preference for massive
neutrinos at more than $2\sigma$. We obtain $\sum m_\nu=0.32 \pm 0.17$ eV combining Planck, BAO and cluster data
and $\sum m_\nu=0.26^{+0.20}_{-0.21}$ eV replacing the latter with shear data. However, looking at the lower panels
of Fig.~\ref{fig:CMB_single_mnu} it is clear that the large mean value of $\sum m_\nu=0.32$ eV obtained from
Planck+BAO+Cluster is driven by the tension between Planck+BAO and cluster constraints. In other words,
the resulting constraints cannot be used to claim a significant detection of the neutrino mass,
but rather they represent a compromise solution between discrepant datasets. Indeed, if we repeat the analysis
marginalizing over the bias the best fit improves by $\Delta \chi^2\simeq11$ -- in this case the Planck+BAO and Planck+Cluster($B_M$) contours overlap 
(see middle panel of Fig.~\ref{fig:mnu_variation})-- and we obtain only a mild preference for massive neutrino 
at $1\sigma$ and an upper limit of $\sum m_\nu<0.28$ eV at $2\sigma$. A very tight
upper limit of $\sum m_\nu=0.14$ eV results instead from the combination of Planck with BAO and Ly-$\alpha$ data
in agreement with the previous results obtained from WMAP9 data. Then we add progressively the shear and Ly-$\alpha$
constraints to the joint analysis Planck+BAO+Cluster($B_M$). Again the inclusion of shear measurements after
cluster data does not alter significantly the error on $\sum m_\nu$, but lowers by $\sim 0.02$ the power
spectrum normalization boosting the total neutrino mass to $\sum m_\nu=0.22^{+0.17}_{-0.18}$ eV, thus providing
a $2\sigma$ evidence for massive neutrinos. 
Instead, the shift to higher $\sigma_8$ value induced by the Ly-$\alpha$ dataset pushes again the mean neutrino mass 
toward lower value and wipes out the neutrino mass detection yielding $\sum m_\nu<0.24$ eV. 

Another possible way to relieve the tension between Planck and clusters data is to marginalize over the lensing
contribution to the temperature power spectrum, parametrized by the parameter $A_{\rm L}$. The Planck Collaboration
reported some anomalies when $A_{\rm L}$ is included in the fit: for a $\Lambda$CDM+$A_{\rm L}$ model they found
$A_{\rm L}=1.22^{+0.25}_{-0.22}$~\citep{Planck_cosm_parameters}, which is at $2\sigma$ from the expected value
of one and $1\sigma$ away from the lensing signal extrapolated from the 4-point function
$A_{\rm L}^{\phi\phi}=0.99^{+0.11}_{-0.10}$. Since Planck constraints on neutrino mass mainly relay
on lensing information (massive neutrinos increase the expansion rate at $z\gtrsim 1$ suppressing clustering
on sub-horizon scales at non-relativistic transition; see e.g.~\citep{Planck_cosm_parameters}) marginalizing over
$A_{\rm L}$ significantly degrades the error on $\sum m_\nu$. Moreover, the preferred value of $A_{\rm L}>1$ shifts by $\sim 1\sigma$
the $\Omega_{\rm m}-\sigma_8$ contours bringing Planck in much better agreement with cluster and shear data (see right panel of Fig.~\ref{fig:mnu_variation}).
The joint analysis Planck($A_{\rm L}$)+Cluster gives $\sum m_\nu<0.41$ eV with an improved best fit with respect to the Planck+Cluster
analysis of $\Delta \chi^2\simeq16$, while Planck($A_{\rm L}$)+BAO yields $\sum m_\nu<0.46$ eV with a mild
preference for massive neutrinos similar to the results obtained in combination with WMAP9.
Combining Planck($A_{\rm L}$) with cluster and BAO yields $\sum m_\nu = 0.35^{+0.15}_{-0.16}$ eV: the tight constraints on
$\sigma_8$ and $\Omega_{\rm m}$ provided by the combination of cluster and BAO data along with the low value of 
power spectrum normalization preferred by the former and large value of the matter density parameter preferred by the latter
require large neutrino masses to bring the two datasets into agreement.
The further inclusion of shear data, whose degeneracy direction between $\sigma_8$ and $\Omega_{\rm m}$ overlaps the one
inferred from Planck($A_{\rm L}$) data, does not change the parameter constraints nor shift their preferred values.
For this analysis we do not consider the Ly-$\alpha$ dataset which exhibits a larger than $2\sigma$ tension
with Planck($A_{\rm L}$)+Cluster and Planck($A_{\rm L}$)+shear results in the $\sigma_8-\Omega_{\rm m}$ plane
due to the large value of the power spectrum normalization favoured by Ly-$\alpha$ forest data.

\begin{figure}
\centering
\includegraphics[width=1.0\textwidth]{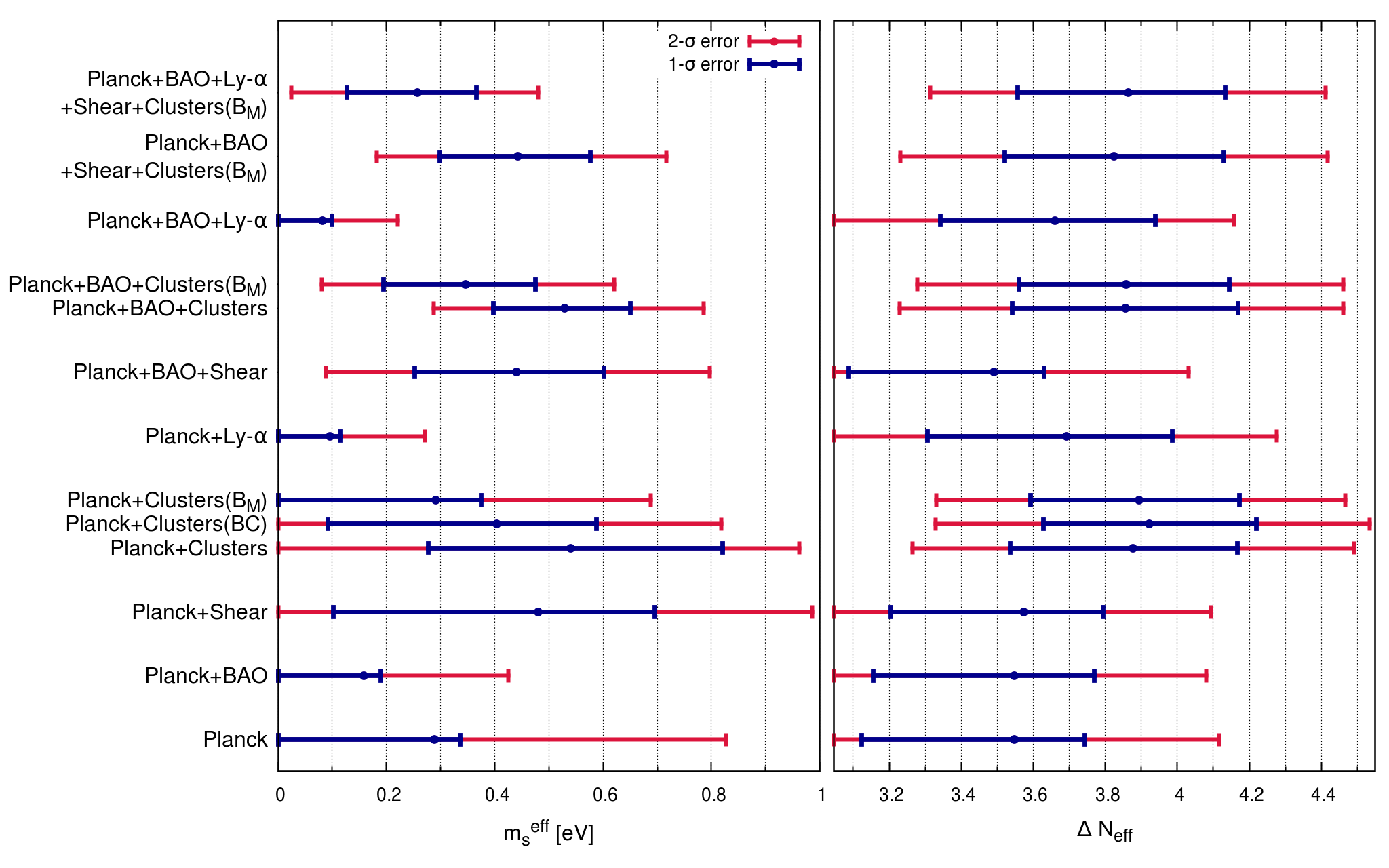}
\caption{Summary of the $1\sigma$ and $2\sigma$ errors on $m_{\rm s}^{\rm eff}$ and $N_\eff$ obtained from the dataset combinations discussed in section~\ref{sub:meff}
for a $\Lambda$CDM+$m_{\rm s}^{\rm eff}$+$N_\eff$ model with one massive neutrino of mass $m_\nu=0.06$ eV.}
\label{fig:meff_nnu_error}
\end{figure}
\begin{table}
    \centering
    %\scriptsize
    \footnotesize
    %\small
    \caption{Constraints on $\Omega_{\rm m}$, $\sigma_8$, $m_{\rm s}^{\rm eff}$ and $\Delta N_{\rm eff}$
    for a $\Lambda$CDM model with massive sterile neutrino using different datasets. Errors are reported
    at $68\%$ confidence level for $\sigma_8$ and $\Omega_{\rm m}$, and both $68\%$ and $95\%$ confidence
    level for $m_{\rm s}^{\rm eff}$ and $\Delta N_{\rm eff}$. Notations included in parenthesis denote
    modifications to the standard setting: (BC) stands for the baryon correction to the HMF, while ($B_M$)
    corresponds to analyses with the mass bias parameter marginalized out.}
    \label{tab:msterile}
    
   \begin{tabular}{lcccccc}
    \hline \vspace{-3mm} \\
	Dataset			&	$\Omega_{\rm m}$	&	$\sigma_8$	&	\multicolumn{2}{c}{$m_{\rm s}^{\rm eff}$[eV]} 	&	\multicolumn{2}{c}{$\Delta N_{\rm eff}$}	\\
				&				&			&	$68\%$CL	&	$95\%$CL	&	$68\%$CL	&	$95\%$CL	\\

   \hline \vspace{-3mm}\\
	Planck			&$0.322^{+0.025}_{-0.030}$& $0.800^{+0.052}_{-0.031}$	& 	$<0.34$		& 	$<0.86$		&$0.49^{+0.18}_{-0.42}$	&	$<1.07$		\vspace{0.5mm} \\
	Planck+Cluster		&$0.304^{+0.026}_{-0.027}$& $0.745^{+0.023}_{-0.037}$	&$0.54^{+0.26}_{-0.26}$& 	$<0.98$		&$0.84^{+0.32}_{-0.32}$	&$0.84^{+0.63}_{-0.60}$\vspace{0.5mm} \\
	Planck+Cluster($B_M$)	&$0.295^{+0.019}_{-0.028}$&$0.794^{+0.040}_{-0.032}$	& 	$<0.38$		& 	$<0.69$		&$0.85^{+0.28}_{-0.30}$&$0.85^{+0.56}_{-0.57}$\vspace{0.5mm} \\
	Planck+Cluster(BC)	&$0.296^{+0.023}_{-0.028}$& $0.770^{+0.031}_{-0.036}$	&$0.40^{+0.31}_{-0.19}$& 	$<0.81$		&$0.88^{+0.30}_{-0.29}$	&$0.88^{+0.61}_{-0.60}$\vspace{0.5mm} \\
	Planck+BAO		&$0.306^{+0.009}_{-0.009}$& $0.818^{+0.033}_{-0.026}$	&	$<0.19$		& 	$<0.43$		&$0.50^{+0.22}_{-0.39}$	&	$<1.04$		\vspace{0.5mm} \\
	Planck+Shear		&$0.309^{+0.028}_{-0.028}$& $0.752^{+0.037}_{-0.043}$	&$0.48^{+0.22}_{-0.38}$& 	$<0.99$		&$0.53^{+0.22}_{-0.37}$	&	$<1.30$		\vspace{0.5mm} \\
	Planck+Ly-$\alpha$	&$0.309^{+0.023}_{-0.024}$& $0.843^{+0.021}_{-0.021}$	& 	$<0.11$		& 	$<0.27$		&$0.65^{+0.30}_{-0.38}$	&	$<1.49$		\vspace{0.5mm} \\
	Planck+BAO+Cluster	&$0.303^{+0.009}_{-0.009}$& $0.744^{+0.013}_{-0.014}$	&$0.53^{+0.12}_{-0.13}$	&$0.53^{+0.26}_{-0.24}$	&$0.81^{+0.31}_{-0.32}$	&$0.81^{+0.60}_{-0.63}$\vspace{0.5mm} \\
Planck+BAO+Cluster($B_M$)	&$0.303^{+0.007}_{-0.009}$& $0.782^{+0.020}_{-0.018}$	&$0.35^{+0.13}_{-0.15}$	&$0.35^{+0.27}_{-0.27}$&$0.81^{+0.29}_{-0.30}$	&$0.81^{+0.60}_{-0.58}$\vspace{0.5mm} \\
	Planck+BAO+Shear	&$0.305^{+0.09}_{-0.010}$& $0.753^{+0.023}_{-0.022}$	&$0.44^{+0.16}_{-0.19}$	&$0.44^{+0.34}_{-0.35}$	&$0.45^{+0.14}_{-0.40}$	&	$<0.99$		\vspace{0.5mm} \\
  Planck+BAO+Ly-$\alpha$	&$0.305^{+0.09}_{-0.010}$& $0.844^{+0.020}_{-0.019}$	&	$<0.10$		& 	$<0.22$		&$0.61^{+0.28}_{-0.32}$	&	$<1.11$		\vspace{0.5mm} \\
Planck+BAO+&\multirow{2}{*}{$0.303^{+0.009}_{-0.009}$}&\multirow{2}{*}{ $0.759^{+0.017}_{-0.020}$}&\multirow{2}{*}{$0.44^{+0.14}_{-0.14}$}&\multirow{2}{*}{$0.44^{+0.28}_{-0.26}$}&\multirow{2}{*}{$0.78^{+0.31}_{-0.30}$}&\multirow{2}{*}{$0.78^{+0.60}_{-0.59}$} \\
	Shear+Cluster($B_M$)	&			&			&			&			&				&			\vspace{0.5mm} \\
Planck+BAO+Ly-$\alpha$&\multirow{2}{*}{$0.293^{+0.009}_{-0.008}$}&\multirow{2}{*}{ $0.794^{+0.016}_{-0.016}$}&\multirow{2}{*}{$0.26^{+0.11}_{-0.13}$}&\multirow{2}{*}{$0.26^{+0.22}_{-0.24}$}&\multirow{2}{*}{$0.82^{+0.27}_{-0.31}$}&\multirow{2}{*}{$0.82^{+0.55}_{-0.55}$} \\
	Shear+Cluster($B_M$)	&			&			&			&			&				&			\vspace{0.5mm} \\
   \hline
    \end{tabular}
\end{table}
\begin{figure}
\centering
\includegraphics[width=0.66\textwidth]{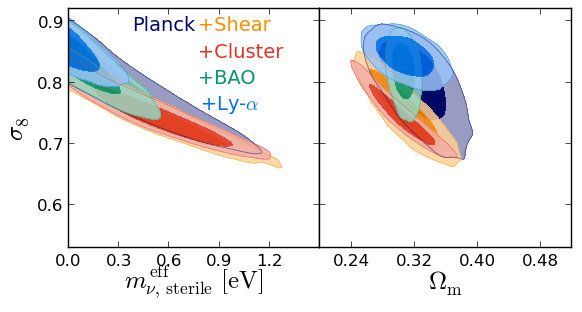}
\includegraphics[width=0.32\textwidth]{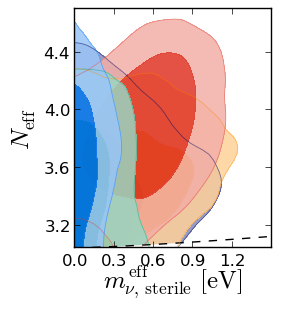}
\caption{Two dimensional likelihood contours at 68\% and 95\% CL for $\sigma_8 - (\Omega_{\rm m},
m_{\rm s}^{\rm eff})$ and $N_{\rm eff} - m_{\rm s}^{\rm eff}$ from Planck combined with different
probes of the low redshift Universe within a $\Lambda$CDM+$m_{\rm s}^{\rm eff}$+$\Delta N_{\rm eff}$ model.
The dashed line shown in the $N_{\rm eff} - m_{\rm s}^{\rm eff}$ plane represent the prior on the
physical sterile neutrino mass: $m_{\rm s}^{\rm eff}/(\Delta N_{\eff})^{3/4} < 10$ eV .}
\label{fig:CMB_single_msterile}
\includegraphics[width=0.36\textwidth]{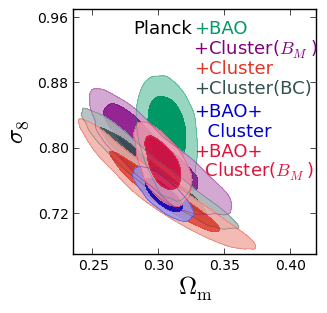}
\includegraphics[width=0.63\textwidth]{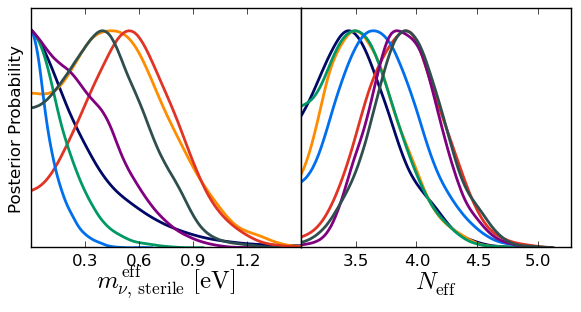}
\caption{\textit{Left} panel - Comparison of the two dimensional likelihood contours at 68\% and 95\% CL in the $\sigma_8 - \Omega_{\rm m}$
plane from the joint analysis of Planck or Planck+BAO and cluster data using different prescriptions for the HMF:
the standard one (\textit{Cluster}), the baryon correction (\textit{Cluster(BC)}) or marginalizing over the bias (\textit{Cluster}($B_M$)).
\textit{Right} panels - Posterior probability distributions for $m_{\rm s}^{\rm eff}$ and $N_{\rm eff}$ from single datasets
combined with Planck within a $\Lambda$CDM+$m_{\rm s}^{\rm eff}$+$\Delta N_{\rm eff}$ model. The colour coding of the lines is the
same of Fig.~\ref{fig:CMB_single_msterile} and the \textit{left} panel of this figure.}
\label{fig:CMB_nnu_msterile_1d}
\end{figure}

\subsection{Extra sterile massive neutrinos}\label{sub:meff}
We now explore the scenario with an extra sterile neutrino component. Table~\ref{tab:msterile} and Fig.~\ref{fig:meff_nnu_error} summarize
the results obtained for the various data combinations. For this cosmological model we employ only Planck as CMB dataset
since the constraints obtained from WMAP9 are much weaker than Planck ones. With the same logic adopted in the previous
section we start combining single dataset with the CMB data and then we add them progressively. The inclusion
of $N_{\rm eff}$ in the fit opens new parameter degeneracies which relax the Planck contours and bring the cluster
and shear constraints in better agreement with the CMB data.
In particular, $N_\eff>3.046$ increases the radiation energy content (see Eq.~\ref{eqn:neff2})
and affects the expansion rate of the Universe thus relaxing the bounds on 
$H_0$ and the scalar spectral index, with which $N_\eff$ is positively correlated. 
When cluster constraints are included we find a mild preference
for massive sterile neutrino $m_{\rm s}^{\rm eff}=0.54\pm0.26$ eV (68\%), and a $2\sigma$ hint for extra
dark radiation $\Delta N_{\rm eff}=0.84^{+0.63}_{-0.60}$. 
At variance with the Planck+Cluster joint analysis performed in the previous section 
-- i.e. within a $\Lambda$CDM+$\sum m_\nu$ model -- this time the Planck+Cluster combination does not
exhibit strong tension with the Planck results, which improves the $\chi^2$ best fit by $\simeq11$ (e.g. compare the $\sigma_8-\Omega_{\rm m}$ panels
of Fig.~\ref{fig:CMB_single_mnu} and Fig.~\ref{fig:CMB_single_msterile}).
%At variance with the Planck+Cluster joint analysis within
%$\Lambda$CDM+$\sum m_\nu$ this time the power spectrum normalization is in  agreement with the Planck results. 
This fact, along with the lower $\sigma_8$ and larger $H_0-n_s$ values preferred by cluster data and the positive correlation between
$m_{\rm s}^{\rm eff}$ and $\Delta N_{\rm eff}$ accounts for the shift of the two parameters with respect
to the Planck-only analysis (see Fig.~\ref{fig:CMB_single_msterile}).
Analogous constraints on $m_{\rm s}^{\rm eff}$ are provided by the inclusion of shear measurements which lowers the $\sigma_8$
mean value, while leaving the bounds on $n_s$ and $H_0$, and thus on $\Delta N_\eff$, unaffected with respect to the Planck-only results.
If we consider the Planck+Cluster analysis, the BC to the HMF results in a $1\sigma$
shift of the power spectrum normalization toward higher values which
reduces the $m_{\rm s}^{\rm eff}$ mean value, while keeping the constraints on $N_\eff$ and the best fit value almost unchanged (see Fig.~\ref{fig:CMB_nnu_msterile_1d}).
Similarly, but with an increased magnitude, if we repeat the analysis marginalizing over the bias the preferred $\sigma_8$ value shifts by $2\sigma$
at the expense of a large value for the bias, $B_M \sim 0.8$, wiping out the former $1\sigma$ preference for $m_{\rm s}^{\rm eff}$ larger than zero.
For this analysis we find a mild improvement of the best fit value of $\Delta \chi^2\simeq5$ with respect to the standard one.
 At odds the inclusion of BAO data reduces
the error on $\Omega_{\rm m}$ and slightly increases the $\sigma_8$ mean value with respect to the Planck-only analysis. This results in tighter constraints
for the sterile neutrino mass, $m_{\rm s}^{\rm eff}<0.43$ eV, and leaves almost unchanged the bounds on $\Delta N_{\rm eff}$. 
When joined to the Planck analysis the Ly-$\alpha$ data constrains $\sigma_8$ in the high values region allowed by Planck data and slightly increases
the $n_{\rm s}$ and $H_0$ mean values. This entails an upper limit of $0.27$ eV for the effective sterile neutrino mass and a $\sim 20\%$ increase of the $\Delta N_\eff$ mean values.

\begin{figure}
\centering
\includegraphics[width=0.66\textwidth]{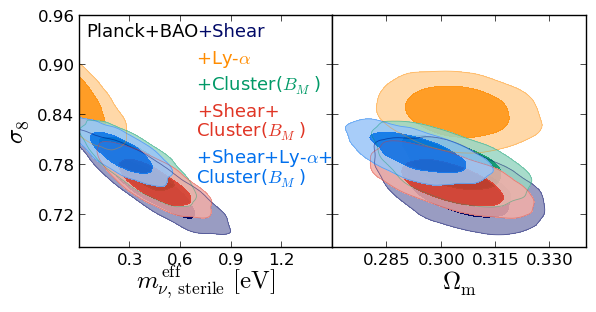}
\includegraphics[width=0.32\textwidth]{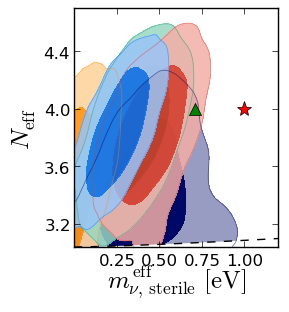}
\includegraphics[width=0.66\textwidth]{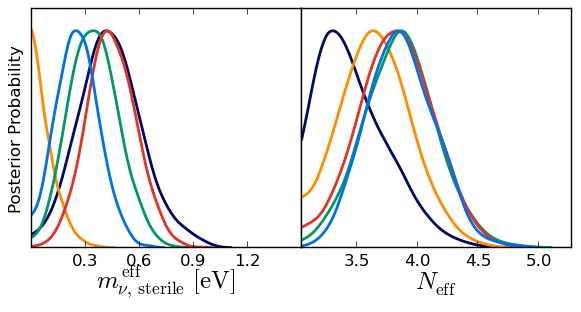}
\caption{\textit{Upper} panels - Confidence contours at 68\% and 95\% CL in the $\sigma_8 - (\Omega_{\rm m},
m_{\rm s}^{\rm eff})$ and $N_{\rm eff} - m_{\rm s}^{\rm eff}$ planes from Planck combined with various
probes of the low redshift Universe within a $\Lambda$CDM+$m_{\rm s}^{\rm eff}$+$\Delta N_{\rm eff}$ model. Also shown
in the $N_{\rm eff} - m_{\rm s}^{\rm eff}$ plane the value of the effective sterile neutrino masses suggested by accelerator (\textit{green triangle})
and reactor and Gallium (\textit{red star}) anomalies assuming $\Delta N_\eff=1$. 
\textit{Lower} panels - 1D likelihood distributions for $m_{\rm s}^{\rm eff}$ and $N_{\rm eff}$ for the same datasets combinations.}
\label{fig:CMB_all_msterile}
\end{figure}
 
We start now to combine Planck CMB measurements with different low redshift probes.
The main results of these analyses are shown in Fig.~\ref{fig:CMB_all_msterile}. Also shown in the $N_{\rm eff} - m_{\rm s}^{\rm eff}$ plane
are the $m_{\rm s}^{\rm eff}$ values motivated by reactor and Gallium anomalies ($m_{\rm s}^{\rm eff} \sim 1$ eV) and accelerator anomaly ($m_{\rm s}^{\rm eff}\sim 0.70$ eV), for
a fully-thermalised sterile neutrino component ($\Delta N_\eff = 1$).
As for the $\Lambda$CDM+$\sum m_\nu$ model,
including BAO data in the Planck+Cluster or Planck+Shear analyses provides a $2\sigma$ evidence for massive sterile neutrino --
$m_{\rm s}^{\rm eff}=0.53^{+0.26}_{-0.24}$ eV and $m_{\rm s}^{\rm eff}=0.44^{+0.34}_{-0.35}$ eV, respectively -- reason
for that being the tight constraints on $\Omega_{\rm m}$ from BAO measurements, which allow to break the $\sigma_8 - \Omega_{\rm m}$ degeneracy,
and the low $\sigma_8$ mean value preferred by cluster and shear data. As for $\Delta N_\eff$ the Planck+BAO+Shear combination shows only
a mild preference for $\Delta N_\eff >0$, while the Planck+BAO+Cluster joint analysis, driven by the large $n_{\rm s}$ value required
by cluster data, favours at $2\sigma$ an extra dark radiation component.
However, even if not as strong as for the $\Lambda$CDM+$\sum m_\nu$ model, the Planck+BAO and Planck+Cluster results still exhibits
a $\sim 1\sigma$ tension (see \textit{right} panel of Fig.~\ref{fig:CMB_nnu_msterile_1d}),
which could drive the large $m_{\rm s}^{\rm eff}$ value obtained from the combination of the three datasets.
Interestingly, at variance with the $\Lambda$CDM+$\sum m_\nu$ analysis, the $2\sigma$ detection of massive sterile neutrino
remains also if we repeat the Planck+BAO+Cluster analysis including the uncertainty in the mass bias. In this case Planck, BAO and
cluster($B_M$) results are in good agreement and the inclusion of
cluster data in the Planck+BAO joint analysis does not shift the contours outside the region allowed by the latter
but only shrinks the errors, as expected for consistent datasets. Compared to the previous analysis without mass bias the fit improves by $\Delta \chi^2\simeq5$. 
The inclusion of $B_M$ in the fit increases the $\sigma_8$ mean value by $\sim 0.04$ which involves a decrease of $\sim 35\%$
of the mean effective sterile neutrino mass, $m_{\rm s}^{\rm eff}=0.35\pm 0.27$ eV, while leaving unaffected the bounds on $\Delta N_\eff$.
This datasets combination rejects at more than $3\sigma$ the vanilla $\Lambda$CDM model, which is located at the origin of 
$N_{\rm eff} - m_{\rm s}^{\rm eff}$ plane in Fig.~\ref{fig:CMB_all_msterile}, but it also disfavours a sterile neutrino mass of $m_{\rm s}^{\rm eff}\sim 0.70$ eV
as suggested by accelerator experiments, and even more strongly rejects the value of $m_{\rm s}^{\rm eff}\sim 1$ eV motivated by reactor and Gallium experiments.
The further inclusion of shear measurements has the effect of favour a slightly lower value of $\sigma_8$, which increases
the mean sterile neutrino mass by $30\%$ with respect to the Planck+BAO+Cluster($B_M$) analysis, thus strengthening
the evidence for a light sterile neutrino species. Moreover, this shift is sufficient to bring the $N_{\rm eff} - m_{\rm s}^{\rm eff}$ confidence
contours in agreement with the sterile neutrino mass motivated by accelerator experiments within $2\sigma$. 
As in the massive active neutrino scenario, the inclusion of Ly-$\alpha$ data on the Planck+BAO analysis leads to opposite results.
The large value of $\sigma_8$ preferred by Ly-$\alpha$ data requires small value for the sterile neutrino mass to be consistent
with the Planck+BAO constraints. This provides a thigh upper limit of $0.22$ eV on $m_{\rm s}^{\rm eff}$ in agreement with the standard $\Lambda$CDM predictions.
Finally, we add the Ly-$\alpha$ datasets to the combination Planck+BAO+Shear+Cluster($B_M$) finding $m_{\rm s}^{\rm eff}=0.26^{+0.22}_{-0.24}$ eV.
The Ly-$\alpha$ contours are in tension with the Planck+BAO+Shear+Cluster($B_M$) results at more than $1\sigma$,
and these constraints on $m_{\rm s}^{\rm eff}$ reflect the compromise between the low $\sigma_8$ mean value
preferred by shear and cluster data and the large one inferred from Ly-$\alpha$ data.
As for $\Delta N_\eff$, the inclusion of the Ly-$\alpha$ data does not shift its mean value but
helps to reduce the errors giving $\Delta N_\eff = 0.82\pm0.55$, that is a larger than $2\sigma$
preference for extra dark radiation.

%%%%%%%%%%%%%%%%%%%%%%%%% conclusions %%%%%%%%%%%%%%%%%%%
\section{Conclusions}
\label{sec:conc}
Many recent studies investigated extensions of the $\Lambda$CDM model that include massive neutrinos
as a possible means to relieve the tension between Planck CMB data and several measurements
of the large scale structure (LSS), finding preferences for non-vanishing neutrino mass.
The neutrino mass claimed detections are mainly driven by low redshift growth of structure constraints,
and in particular by galaxy clusters data, thanks to the tight constraints that they provide on the combination $\sigma_8(\Omega_{\rm m}/0.27)^\gamma$.
However, neutrino constraints from cluster data suffer from systematic errors manly
related to the uncertainty in the bias of the mass-observable relation and the calibration
of the HMF in cosmology with massive neutrino, which often are not taken into account in data analysis.
Here we presented neutrino mass constraints, either for an active and sterile neutrino scenario,
from the combination of CMB measurements with low redshift Universe probes. We employed for our
analysis CMB measurements from WMAP9 or Planck in combination with BAO scale measurements from
BOSS DR11, galaxy shear power spectrum from CFTHLenS, SDSS Ly-$\alpha$ forest power spectrum
constraints and cluster mass functions from \textit{Chandra} observations. 
At variance with previous similar studies, which included in their 
analysis constraints derived within a vanilla $\Lambda$CDM model,
%Instead of including constraints derived for a vanilla $\Lambda$CDM model
we performed a full likelihood analysis for all the datasets employed in this work
in order to avoid model dependence of the constraints. Moreover, in
the cluster data analysis we properly take into account 
the impact of massive neutrino in the HMF calibration and we investigated the effects
on cosmological constraints of the uncertainty in the mass bias and re-calibration 
of the HMF due to baryonic feedback processes as suggested in~\citep{Cui2014}. 

For both neutrino scenarios assumed and CMB datasets employed, we
found that none of the constraints from the LSS data, combined on a
one-by-one basis, with CMB measurements provide strong -- i.e. larger
than $2\sigma$ -- evidence for massive neutrino.  From the joint
analysis Planck+Cluster we obtained $\sum m_\nu <0.34$ eV but we
emphasize that the extensions to three massive active neutrinos is not
sufficient to bring the dataset in agreement with Planck results.
Indeed the extension to massive neutrinos does not improve the
fit of the combination of Planck and galaxy cluster data, with
respect to the vanilla $\Lambda$CDM model.  Taking into account the
effect of baryons on the HMF calibration or the uncertainty in the
mass bias increases the $\sigma_8$ mean value improving the fit with
respect to the standard analysis of $\Delta \chi^2 \simeq 2$ and
$\Delta \chi^2 \simeq 9$, respectively. In the latter case
  constraints from Planck CMB and galaxy clusters agree within
  $1\sigma$, with their combination preferring a vanishing neutrino
  mass.  Alternatively, the Planck and cluster datasets can be
brought in agreement marginalizing over the lensing contribution to
the temperature power spectrum.  In this case the Planck' s
$\sigma_8-\Omega_{\rm m}$ contours relax and shift by $\sim 1\sigma$,
improving the $\chi^2$ best fit value by $\sim 16$ with respect to the
Planck+Cluster analysis.  Similarly, assuming an extra sterile
neutrino species, which introduces in the fit the additional parameter
$\Delta N_\eff$, relaxes the Planck's bounds reducing the discrepancy
with cluster results by $\Delta \chi^2 \simeq 11$ with respect to the
same data combination in a $\Lambda$CDM+$\sum m_\nu$ model. From this
analysis we obtained a $1\sigma$ preference for non vanishing neutrino
mass and $\Delta N_\eff=0.84^{+0.63}_{-0.60}$.  Including also the
mass bias in the fit further improves the agreement between Planck and
cluster datasets by $\Delta \chi^2\simeq5$ at the expense of an higher
$\sigma_8$ mean value which cancels the former $1\sigma$ detection of
massive neutrinos.  Preference for non-vanishing neutrino mass at more
than $2\sigma$ were found instead combining CMB and BAO measurements
with shear or cluster data.  The BAO constraints break the
$\sigma_8-\Omega_{\rm m}$ degeneracy typical of cluster and shear data
while the low $\sigma_8$ mean value preferred by the latter is
compensated by large neutrino mass.  However, for the
$\Lambda$CDM+$\sum m_\nu$ model, the large neutrino mass obtained from
the joint analysis Planck+BAO+Cluster is driven by the tension between
Planck+BAO and cluster constraints.  Indeed, including the mass bias
parameter in the fit reduces the $\sum m_\nu$ mean value by $50\%$
wiping out the $2\sigma$ preference for massive neutrino, but
increasing the $\chi^2$ best fit value by $\sim 11$.  For the sterile
neutrino case, when considering the uncertainty in the mass bias, the
fit is improved by $\Delta \chi^2 \simeq 5$ at the expenses of a lower
mean neutrino mass but still with a preference for an extra massive
neutrino, $m_{\rm s}^{\rm eff}=0.35\pm0.27 {\rm eV}- \Delta
N_\eff=0.81^{+0.60}_{-0.58}$.  The significance of the detection
increases further including simultaneously shear and cluster data.
For a $\Lambda$CDM model with three degenerate massive neutrinos we
obtained $\sum m_\nu=0.29^{+0.18}_{-0.21}$ eV from the combination
WMAP9+BAO+Shear+Cluster($B_M$), while replacing WMAP9 with Planck
measurements we got $\sum m_\nu=0.22^{+0.17}_{-0.18}$ eV, or $\sum
m_\nu=0.35^{+0.15}_{-0.16}$ eV marginalizing over the lensing signal.
For the sterile neutrino case, from the combination
Planck+BAO+Shear+Cluster($B_M$), we found $m_{\rm s}^{\rm
  eff}=0.44^{+0.28}_{-0.26} {\rm eV}$ and $\Delta
N_\eff=0.78^{+0.60}_{-0.59}$, that is a larger than $3\sigma$
rejection of the vanilla $\Lambda$CDM model. Assuming a fully
thermalised sterile neutrino these constraints reject at even higher
significance a $1.0$ eV sterile neutrino as motived by reactor and
Gallium anomalies, while a neutrino mass of $0.7$ eV as suggested by
accelerator anomaly is within the $2\sigma$ errors.  Conversely, the
Ly-$\alpha$ measurements tend to increase the $\sigma_8$ mean value
with respect to the CMB data analyses, which in turn entails a
preference for vanishing neutrino masses to be consistent with the
other parameters constraints. For the active neutrino scenario we got
$\sum m_\nu<0.19$ eV and $\sum m_\nu<0.14$ eV combining BAO,
Ly-$\alpha$ and WMAP9 or Planck dataset, respectively. Similarly, for
the sterile neutrino model we obtained $m_{\rm s}^{\rm eff}<0.22{\rm
  eV}$ and $\Delta N_\eff<1.11$.  The full data combination provides
neutrino mass constraints which reflects the compromise between the
$\sigma_8$ values preferred by shear and cluster data and those
inferred from Ly-$\alpha$ measurements. For the $\Lambda$CDM+$\sum
m_\nu$ model we obtained only an upper limit on the total neutrino
mass independently from the CMB dataset employed, while in the sterile
neutrino scenario we still found a $2\sigma$ preference for an extra
massive species, $m_{\rm s}^{\rm eff}=0.26^{+0.22}_{-0.24} {\rm eV}-
\Delta N_\eff=0.82\pm0.55$.  

In summary, our results highlight
that current CMB and LSS probes point towards a significant
detection of the sterile neutrino mass and dark radiation unless the
constraints on $\sigma_8$ provided by clusters and shear data turn
out to be biased toward lower values. As for clusters, this could be due
to a possible underestimate of the cluster mass bias. As for cosmic
shear, an underestimate of $\sigma_8$ could be induced 
by a misinterpretation of the intrinsic alignment
signal. On the other hand new BOSS results from the 1D Ly-$\alpha$
flux power spectrum or Planck CMB data -- e.g. due to a different
foreground removal technique~\citep{2013arXiv1312.3313S} -- could
strengthen or weaken the evidence for non-vanishing neutrino
masses.

More in general, our results highlight that current
cosmological data already have the potential to set rather stringent
constraints on neutrino masses, which could even challenge the
results from laboratory experiments, but these are hampered by
systematics which need to be better controlled and understood.  This
becomes even more important in view of future surveys (eROSITA,
SPT3G, DES, DESI, Euclid), that thanks to the large amount of data
to be provided will bring down the statistical errors by large
factors.
As for cluster cosmology, the ever increasing number of high quality
weak lensing data is expected to provide in the near term well
characterized and unbiased constraints on the absolute cluster mass
calibration. From the theoretical side, refined cosmological
simulations which properly accounts for neutrino and baryonic physics
will be crucial to improve the calibration of the HMF and modelling of
the shear and Ly-$\alpha$ forest flux power spectrum.

\let\thefootnote\relax\footnote{Note added: during the finalization of
  this work, Ref~\citep{2014arXiv1407.4516M} presented constrains on
  $\sum m_\nu$ from the combination of WMAP9 or Planck CMB data with a
  sample of 224 cluster from the ROSAT All-Sky survey clusters with
  masses calibrated using weak lensing data from Subaru and CFHT
  telescopes for 50 galaxy clusters.  Even if their results are not
  directly comparable with ours they are qualitatively in agreement.}
%%%%%%%%%%%%%%%%%%%%%%%%%% Acknowledgements %%%%%%%%%%%%%%%
\section*{Acknowledgments} 
We thank Martin Kilbinger and Alexey Vikhlinin for useful discussions.
This work has been supported by the PRIN-MIUR 201278X4FL grant, by the  IS PD51 INDARK INFN grant
and by "Consorzio per la Fisica di Trieste". MV is supported by the ERC Starting Grant CosmoIGM.
%%%%%%%%%%%%%%%%%%%%%%%%% bibliography %%%%%%%%%%%%%%%%%%%

\bibliographystyle{JHEPb}
\bibliography{Bibliography}

\end{document}